\documentclass[pre,aps,reprint,twocolumn,floatfix]{revtex4}

\usepackage{amsmath}
\usepackage{amssymb}
\usepackage{graphicx} 
\usepackage{dcolumn}  
\usepackage{bm}       
\usepackage{subfigure}
\usepackage{natbib}
\usepackage{hyperref}
\usepackage{float}
\usepackage[normalem]{ulem}
\hypersetup{colorlinks=true, pdfstartview=FitV, linkcolor=blue, citecolor=black, plainpages=false, pdfpagelabels=true, urlcolor=blue}
\usepackage[all]{hypcap}

\begin{document}

\title{Transverse Temperature Interfaces in the Katz--Lebowitz--Spohn Driven Lattice Gas}

\author{Ruslan I. Mukhamadiarov} \email{mruslani@vt.edu}
\author{Priyanka} \email{pri2oct@vt.edu}
\author{Uwe C. T\"auber} \email{tauber@vt.edu}
\affiliation{Department of Physics (MC 0435) and Center for Soft Matter and Biological Physics,
		Virginia Tech, Robeson Hall, 850 West Campus Drive, Blacksburg, Virginia 24061, USA}

\date{\today}


\begin{abstract}
We explore the intriguing spatial patterns that emerge in a two-dimensional spatially inhomogeneous Katz--Lebowitz--Spohn (KLS) driven lattice gas with attractive nearest-neighbor interactions. 
The domain is split into two regions with hopping rates governed by different temperatures $T > T_c$ and $T_c$, respectively, where $T_c$ indicates the critical temperature for phase ordering, and with the temperature boundaries oriented perpendicular to the drive. 
In the hotter region, the system behaves like the (totally) asymmetric exclusion processes (TASEP), and experiences particle blockage in front of the interface to the critical region. 
To explain this particle density accumulation near the interface, we have measured the steady-state current in the KLS model at $T > T_c$ and found it to decay as $1/T$. 
In analogy with TASEP systems containing ``slow'' bonds, we argue that transport in the high-temperature subsystem is impeded by the lower current in the cooler region, which tends to set the global stationary particle current value. 
This blockage is induced by the extended particle clusters, growing logarithmically with system size, in the critical region.
We observe the density profiles in both high-and low-temperature subsystems to be similar to the well-characterized coexistence and maximal-current phases in (T)ASEP models with open boundary conditions, which are respectively governed by hyperbolic and trigonometric tangent functions.
Yet if the lower temperature is set to $T_c$, we detect marked fluctuation corrections to the mean-field density profiles, e.g., the corresponding critical KLS power law density decay near the interfaces into the cooler region.
\end{abstract}


\maketitle

\section{\label{sec:level1}Introduction}

The absence of a unified theoretical framework for non-equilibrium systems has instigated the development of a variety of simple models that capture certain decisive features of non-equilibrium processes. 
Driven lattice gases represent a class of paradigmatic interacting particle systems that has attracted considerable attention over the past decades \cite{SchmittmannBook:1995,Schutz:2001,Schmittmann:1998,Marro:1999}. 
These driven lattice gas models are characterized by non-trivial stationary states that display generic scale invariance \cite{Grinstein:1991,Hohenberg:1993,TauberBook:2014,TauberRev:2017}, implying that their dynamics is governed (asymptotically) by power laws and genuine non-equilibrium scaling exponents. 
Prototypical examples are the asymmetric and totally asymmetric exclusion processes (T)ASEP with hard-core particles and (fully) biased hopping transport \cite{Spitzer:1970,Liggett:1985,Derrida:1993a,Johansson:2000,Stinchcombe:2001}. 
Adding attractive nearest-neighbor Ising interactions as in the Katz--Lebowitz--Spohn (KLS) model moreover induces a continuous phase transition at a critical temperature $T_c$ separating disordered configurations from a low-temperature regime showing phase segregation in the form of ordered particle (or hole) stripes oriented parallel to the drive \cite{Katz:1983,Katz:1984,Marro:1999,Schmittmann:1998}. 

The (T)ASEP model and KLS system in the disordered high-temperature phase are governed by identical scaling exponents \cite{Schmittmann:1995,SchmittmannBook:1995}. 
However, the KLS dynamical scaling properties change when the temperature approaches the critical value $T_c$: 
In this critical region, the dynamics is impeded significantly by the diverging correlation length of the order parameter fluctuations, i.e., critical slowing-down. 
Consequently, the asymptotic non-equilibrium scaling exponents assume critical values that are distinct from those that capture generic dynamical scale invariance at elevated temperatures. 

Both in its high-temperature phase and at the non-equilibrium critical point, the KLS model thus displays dynamic scale invariance. 
Hence it is interesting to wonder if a spatially inhomogeneous system with different regions held at different temperatures $T_h > T_c$ and $T_l = T_c$, respectively, remains scale-invariant, and how either spatial domain might influence or even control the transient and stationary properties of the other region. 
Moreover, one may ask how its properties compare to those of the standard homogeneous KLS system at uniform temperature, and if there emerge pronounced boundary effects or even interface-driven spontaneous pattern formation. 
To our knowledge, the static and dynamic features induced by the interfaces generated by combining two KLS driven lattice gases held at distinct temperatures have not yet been explored in the literature. 
In this present work, we set one part of the system at high temperature $T_h > T_c$, whereas the other part is (usually) held at the critical point $T_c$, with the temperature boundaries oriented transverse to the drive.
While the original KLS model \cite{Katz:1983,Katz:1984} was motivated by transport in fast ionic conductors, it is conceivable that spatially inhomogeneous continuum driven systems similar to our two-temperature KLS model variant might be experimentally realized in suspensions of charged colloids with screened Coulomb interactions driven by external electric fields.

The effects of similar temperature heterogeneities were earlier explored in detailed studies of an Ising ring (without external drive) \cite{Pleimling:2014} and two-dimensional square lattice \cite{Li:2012}. 
Other two-temperature lattice gas models were introduced by associating different temperatures $T_x$, $T_y$ with different hop directions $x, y$ \cite{Praest:2000}, or by coupling the $i + j =$ even / odd sublattices to different temperature reservoirs \cite{Dickman:2016}. 
The key difference between the KLS driven lattice gas and previously studied models resides in the way detailed balance is violated. 
In the two-temperature non-driven Ising models, the second thermal reservoir induces a net heat flux that forces the system out of equilibrium. 
In contrast, the externally imposed hopping bias or drive in the KLS system generates a global net particle current, and consequently a quite distinct non-equilibrium steady state. 
Moreover, the drive induces a strong anisotropy that causes correlations to scale differently in directions parallel and transverse to the drive, with associated distinct correlation lengths $\xi_\parallel \sim \xi_\perp^{1+\Delta}$ with an anisotropy exponent $\Delta$, which in turn leads to spatially anisotropic scaling laws.

Our two-temperature generalization of the driven KLS lattice gas can be considered as two subsystems on a two-dimensional torus coupled to each other through particle exchange across the two interfaces. 
Other interacting KLS subsystems have been investigated, such as a multi-layered KLS model with particle-hole exchange between layers first excluding Ising interactions between layers \cite{Marro:1995}, and subsequently accounting for exchange energetics \cite{Hill:1996}.
In the latter case, the presence of inter-layer interactions produces intriguing spatial patterns termed ``fingers'' or ``icicles''. Multiple-species KLS variants in two dimensions were, e.g., analyzed in Refs.~\cite{Aertsens:1990,Korniss:1997}, where a set of microscopic dynamical rules for two distinct particle species A and B were imposed that resulted in ordering perpendicular to the drive. 
For a more comprehensive list of KLS model variations we refer to the overview~\cite{Schmittmann:1998}.

In the stationary state, the two-temperature KLS model surprisingly displays phase separation into low- and high-density regions in the hotter region, with the separating domain wall located in the middle of the hot subsystem and oriented perpendicular to the drive. 
The high-density phase is formed at the hot-into-critical temperature interface; it is caused by a blockage of particle flow into the critical subsystem, which is impeded by the emerging extended critical clusters.
This in turn generates an algebraic density decay inside the critical region near the boundaries to the hotter subsystem. 
The density profiles of the two subsystems in the two-temperature KLS resemble the standard hyperbolic and trigonometric tangent functions that are characteristic signatures of the coexistence and maximal-current phases in (T)ASEP systems with open boundaries. 
This similarity allows us to analyze the stationary-state properties of the two-temperature KLS model explicitly. 
Employing numerical simulations and mean-field calculations for the open TASEP coupled to particle reservoirs, we find that the uniform steady-state current is set by the impeded transport in the low-temperature region.
Moreover, we detect enhanced spatial fluctuations and marked deviations from the mean-field predictions in the stationary density profile inside both the hotter and cooler regions that are likely induced by emergent long-range correlations emerging from the critical subsystem.
We emphasize that the dynamical interaction of the two KLS systems held at different temperatures through transport across the two interfaces aligned transverse to the external drive and net overall current therefore does note induce mere short-range boundary effects as would be expected in similarly coupled systems near thermal equilibrium. 
In contrast, the cooler region causes spontaneous spatial pattern formation in the hotter subsystem and thereby destroys generic scale invariance there.
In addition, if the colder domain is held at the KLS critical temperature, it imprints the associated strong fluctuations onto temporal correlations that characterize the domain wall separating the high- and low-density phases in the hot region.

The outline of this paper is as follows: 
In the following Sec.~\ref{sec:level2}, we present the microscopic dynamical rules defining the KLS and (T)ASEP models, and describe their continuous coarse-grained description in terms of non-linear Langevin equations. 
We also introduce their scaling properties in the stationary state as well as in the physical aging scaling regime. 
In Sec.~\ref{sec:level3}, we introduce our two-temperature KLS model with the temperature interfaces oriented perpendicular to the drive, and explain our Monte Carlo simulation algorithm. 
We then provide our numerical data and characterize the transient kinetics as well as the stationary current and density profiles in detail in Secs.~\ref{sec:level4} and \ref{sec:level5}. 
We summarize of results and discuss a few open questions in our concluding Sec.~\ref{sec:level6}.

\section{\label{sec:level2}Driven Lattice Gases}
\subsection{\label{sec:sublevel21}KLS Model Description}

The Katz--Lebowitz--Spohn (KLS) model comprises a collection of $N$ (classical) binary variables, either spin up / down $s_i = \pm 1$ or particle occupation numbers $n_i = \tfrac12 (s_i + 1) = 1, 0$, on a $d$-dimensional lattice with $L_\parallel \times L_\perp^{d-1}$ sites, subject to nearest-neighbor attractive Ising interactions and periodic boundary conditions.
A spin or particle may exchange its position with any of its nearest neighbors with a fixed rate depending on the temperature and a uniform external drive field $E \geq 0$ that is oriented along the longitudinal ($\parallel$) direction \cite{Katz:1983, Katz:1984}. 
Spin exchange processes hence are biased along the direction of the applied field, and in conjunction with the periodic boundaries drive the system towards a genuine non-equilibrium stationary state. 
For such a steady state to be maintained in the presence of the external driving field, the system needs to be coupled to a thermal reservoir at fixed temperature $T$ that absorbs heat flow produced by the current. 
Similarly to the Ising model, nearest-neighbor ferromagnetic interactions spur competition in the system between two processes: local ordering or domain wall annihilation, and domain wall formation accompanied with entropy production. 
As a result, the KLS driven lattice gas experiences a continuous phase separation transition in $d \geq 2$ dimensions when the conserved total magnetization of the system is set to vanish, $\sum_{i=1}^N s_i = 0$, or the particle density $\rho = \sum_{i=1}^N n_i / N = \tfrac12$. 
In contrast to the equilibrium Ising model, the drive in the KLS system forces ordered domains to align into stripe-like clusters that are oriented along the $\parallel$ direction. 
It also raises the critical temperature, e.g., $T_c^{\rm KLS}(E \to \infty) \approx 1.41 \, T_c^{\rm eq}$ in two dimensions \cite{Marro:1999}; and as discussed below, drastically alters the associated critical exponents.

The dynamics of the KLS model in $d = 2$ dimensions can be described in the lattice gas language by the following set of microscopic rules: 
Consider a half-filled driven lattice gas that is defined on a torus with $L_\parallel \times L_\perp$ sites, with each site containing at most one particle, restricting the occupation number $n_i = n(x,y) = 0$ (empty site) or $1$ (filled site). 
This constraint on the particle occupation number may be interpreted to reflect mutual hard-core repulsion. 
In addition, the driven KLS lattice gas is governed by nearest-neighbor attractive Ising interactions with uniform exchange coupling $\mathcal{J} > 0$,
\begin{equation}\label{eqn:IsingHamilt}
	H = - \mathcal{J} \sum_{i\not=j}^N s_i \, s_j = - 4 \mathcal{J} \sum_{i\not=j}^N \left( n_i - \tfrac12 \right) \left( n_j - \tfrac12 \right) \ . 
\end{equation}
The drive and temperature enter the model dynamics through the Markovian transition rates 
\begin{equation}\label{eqn:KLSrates}
	R(\mathcal{C} \to \mathcal{C'}) \propto \exp \left(- \beta \, [H(\mathcal{C'}) - H(\mathcal{C}) - l E] \right) \ ,
\end{equation} 
where $\beta = 1 / k_{\rm B} T$, $\mathcal{C}$ and $\mathcal{C'}$ denote two distinct system configurations $\{ s_i \}$ or $\{ n_i \}$, $E > 0$ is the applied drive strength, and $l = {+1,0,-1}$ respectively indicates hops along, transverse to, and against the external bias. 
For $E = 0$, detailed balance is restored and one recovers the Kawasaki exchange dynamics on the Ising lattice. 
As $E \to \infty$, particle motion or spin exchanges in the $\parallel$ direction cannot proceed against the drive, effectively restricting the transition options to $l = 1, 0$ (totally biased case). 
In the limit $T \to \infty$ ($\beta \to 0$), the nearest-neighbor Ising interactions become ineffective, and the KLS model reduces to the asymmetric exclusion process (ASEP) for finite driving field $E$, or the totally asymmetric exclusion process (TASEP) when $E \to \infty$.

\subsection{\label{sec:sublevel22}(T)ASEP Model Description}
The (totally) asymmetric exclusion process (T)ASEP represents one of the paradigmatic models of non-equilibrium dynamics, as the underlying conservation law and the hard-core repulsion between particles induce spatially anisotropic generic scale invariance with non-trivial (i.e., non-diffusive) scaling exponents.
Remarkably, exact solutions are available for the (T)ASEP in one dimension \cite{Derrida:1993a, Derrida:1992,Derrida:1993b, Schutz:1993}.

(T)ASEP dynamics consists of (fully) biased particle diffusion on a lattice subject to site exclusion and periodic boundary conditions; the term ``exclusion'' again refers to the occupation number restriction $n_i = 0$ or $1$. 
Provided the target lattice site is empty, the transition probabilities are set to $p_\parallel > q_\parallel$ and $p_\perp = q_\perp$ for hops along, against, and transverse to the drive direction, respectively. 
The hopping rate difference $\propto p_\parallel - q_\parallel$ in the $\parallel$ direction produces a non-zero probability current parallel to the drive that in combination with the periodic boundary conditions forces the system out of equilibrium.
Henceforth we set both the lattice spacing and the microscopic hopping time step duration to unity. 
For a system with uniform density one then obtains the mean particle current $\langle J_\parallel \rangle = (p_\parallel - q_\parallel) \, \rho (1 - \rho)$.
In this non-equilibrium steady state, the microstates in the periodic (T)ASEP occur with equal probability \cite{Meakin:1986} that is just given by the inverse of the total number of possible states for the $N$ particles to be distributed on a lattice with $V = L_\parallel L_\perp^{d-1}$ sites, $P(V,N) = N! (V - N)! / V!$. 

A much richer phenomenology emerges in the one-dimensional TASEP ($q_\parallel = 0$) with open boundary conditions where particles are injected into the system from a reservoir on one end with injection probability $\alpha \in [0,1]$, and removed on the opposite boundary with ejection probability $\beta \in [0,1]$. 
The non-equilibrium steady states in TASEP systems with such open boundaries depend on the values of those injection / ejection rates and can be classified into four distinct phases \cite{Schutz:1993, Krug:1991, Schutz:2001, Blythe:2007, Stinchcombe:2011}:
\begin{itemize}
\item[$\bullet$] When $\beta < \text{min}(\alpha,\tfrac12)$, the system is in the {\em high-density phase} with bulk density $\rho = 1 - \beta$ and average current $\langle J \rangle = \beta (1 - \beta)$.
\item[$\bullet$] For $\alpha < \text{min}(\beta,\tfrac12)$, the system resides in the {\em low-density phase} with bulk density $\rho = \alpha$ and average current $\langle J \rangle = \alpha (1 - \alpha)$.
\item[$\bullet$] When $\alpha = \beta < \tfrac12$, the system is in the {\em coexistence phase}. 
	The particle density in this ``mixed-state'' region is inhomogeneous and in the stationary state follows a shock profile interpolating between the low- and high-density values $\alpha$ and $1 - \alpha$. 
	A mean-field calculation predicts a hyperbolic tangent density profile 
\begin{equation}\label{eqn:TASEPtanh}
\rho(x) = \tfrac12 \left( 1 + k \tanh \left[ k (x - x_0) \right] \right) ,
\end{equation}
	where the inverse length scale $k = 1 - 2 \alpha = \sqrt{1 - \langle J \rangle / J_c}$ defines both the shock height and width; here $J_c = p_\parallel / 4$ denotes the critical current.
\item[$\bullet$] When both $\alpha, \beta \geq \tfrac12$, the system is in the {\em maximal current phase}.
	In this region the mean steady-state current $\langle J \rangle$ exceeds the critical current $J_c$. 
	The mean-field calculation now yields a trigonometric tangent density profile
\begin{equation}\label{eqn:TASEPtan}
\rho(x) = \tfrac12 \left( 1 - q \tan \left[ q (x - x_0) \right] \right) ,
\end{equation}
	with $q = \sqrt{\langle J \rangle / J_c - 1}$. 
\end{itemize}
In the high- or low-density phase, the density profile decays exponentially with finite characteristic length $k^{-1}$.
However, for the maximal-current state, the characteristic length $q^{-1}$ diverges as $J \to J_c$.

Another one-dimensional model variation relevant to this work is a TASEP lattice gas with slow bonds, or inhomogeneous TASEP \cite{Janowsky:1994}. 
In that case, particles hop across normal bonds with rate $1$, but across slow bonds with the reduced rate $r < 1$. 
Remarkably, just introducing a single slow bond with $r < 1$ in the TASEP results in particle blockage at the position of this defect, and forces the steady-state current to decrease. 
The density of the particles before and after the slow bond $\rho_\pm$ depends on the maximum stationary current $\langle J(r) \rangle$ that the system can sustain:
$\rho_\pm = \tfrac12 \left( 1 \pm \sqrt{1 - \langle J(r) \rangle / J_c} \right)$. 
An explicit expression for the maximal stationary current $\langle J(r) \rangle$ has been obtained by means of a series expansion for $r \lessapprox 1$ in Ref.~\cite{Lebowitz:2013}; a general result for arbitrary values of $r$ remains yet to be found. 

\subsection{\label{sec:sublevel23}Coarse-Grained Langevin Representation}
For the driven KLS lattice gas no exact solution has been found to date.
Yet one may formulate a coarse-grained mesoscopic continuous description, which subsequently allows for a thorough analysis by means of the dynamic renormalization group and determination of the associated scaling exponents.
The starting point is the continuity equation for the conserved order parameter field
\begin{equation} \label{eqn:Continuity}
\frac{\partial s(\vec{x},t)}{\partial t} = - \vec{\nabla}\cdot \vec{J}(\vec{x}, t), 
\end{equation}
where $s(\vec{x},t) = 2 [\rho(\vec{x}, t) - \langle \rho \rangle]$ represents the local magnetization density or density deviation from its mean $\langle \rho \rangle = \tfrac12$. 
The conserved current density comprises three contributions, namely a relaxational term with Onsager coefficient $D$, a non-linear component proportional to the drive $\vec{\mathcal{E}} \rho (1 - \rho) = \tfrac12 D g (1 - s^2) {\hat e}_\parallel$ that also incorporates the exclusion constraint, and Gaussian white noise $\vec{\eta}$ that introduces stochasticity in the system \cite{SchmittmannBook:1995, Leung:1986, Janssen:1986}:
\begin{equation} \label{eqn:current}
\vec{J}(\vec{x},t) = - D \vec{\nabla} \frac{\delta \mathcal{H}[s]}{\delta s(\vec{x},t)} + \frac{\vec{\mathcal{E}}}{4} \left[ 1 - s(\vec{x},t)^2 \right] + \vec{\eta}(\vec{x},t) ,
\end{equation}
Since we are interested in the behavior near the continuum phase transition, we here employ the standard Landau--Ginzburg--Wilson Hamiltonian for a scalar order parameter \cite{Amit:1984}
\begin{equation} \label{Hamiltonian}
\mathcal{H}[s] = \int d^d x \left( \frac12 \left[ \vec{\nabla} s(\vec{x}) \right]^2 + \frac{\tau}{2} s(\vec{x})^2 + \frac{u}{4!} s(\vec{x})^4 \right) ,
\end{equation}
where $\tau \propto |T - T_c|$ denotes the distance from the critical point, and $u > 0$ is the non-linear coupling driving the phase transition.

Taking the functional derivative of the Hamiltonian \eqref{Hamiltonian} and substituting Eq.~\eqref{eqn:current} back into the continuity equation yields a Langevin equation that models the KLS dynamics. 
Near the critical point, fluctuations only need to be taken into account in the ``soft'' ($d-1$)-dimensional transverse (``$\perp$'') spatial sector, since below $T_c$ ordered stripes form only in alignment with the (``$\parallel$'') drive direction. 
The critical KLS Langevin equation hence becomes (after straightforward rescaling):
\begin{multline} \label{eqn:LEQ}
\frac{\partial s(\vec{x},t)}{\partial t} = D \Bigl[ c \nabla_\parallel^2 s(\vec{x},t) + \left( \tau - \nabla_\perp^2 \right) \nabla_\perp^2 s(\vec{x},t) \\
+ \frac{u}{6} \nabla_{\perp}^2 s(\vec{x},t)^3 + \frac{g}{2} \nabla_\parallel s(\vec{x},t)^2 \Bigr] + \sigma(\vec{x},t) ,
\end{multline}
where $D c$ and $D \tau$ now represent longitudinal and transverse diffusion coefficients, respectively, and $\sigma = - {\vec \nabla} \cdot \vec{\eta}$ constitutes conserved Gaussian white noise with zero mean $\langle \sigma \rangle = 0$ and correlations $\langle \sigma(\vec{x},t) \sigma(\vec{x}',t') \rangle = - 2 D \nabla_\perp^2 \delta(\vec{x}-\vec{x}') \delta(t-t')$. 

In a fully analagous manner, one may construct a coarse-grained mesoscopic description for the (T)ASEP; indeed, we merely need to omit the terms in Eq.~\eqref{eqn:LEQ} for the KLS model that pertain to critical fluctuations near $T_c$, and arrive at (again, after straightforward rescaling)
\begin{multline} \label{eqn:TASEPLangevin}
\frac{\partial s(\vec{x},t)}{\partial t} = D \left[ \left( c \nabla_\parallel^2 + \nabla_\perp^2 \right) s(\vec{x},t) + \frac{g}{2} \nabla_\parallel s(\vec{x},t)^2 \right] \\
+ \sigma(\vec{x},t) ,
\end{multline}
with conserved Gaussian white noise $\langle \sigma \rangle = 0$ and $\langle \sigma(\vec{x},t) \sigma(\vec{x}',t') \rangle = - 2 D \left( {\tilde c} \nabla_\parallel^2 + \nabla_\perp^2 \right) \delta(\vec{x}-\vec{x}') \delta(t-t')$, where the ratio $w = {\tilde c} / c$ indicates the deviation from thermal equilibrium, since for $w = 1$ Einstein's relation and detailed balance are satisfied. 

\subsection{\label{sec:sublevel24}Stationary Scaling Exponents}
Upon approaching the phase transition, the static and dynamical correlations described by Eq.~\eqref{eqn:LEQ} become strongly anisotropic, with 
longitudinal and transverse wave vectors scaling as $|q_\parallel| \sim |{\vec q}_\perp|^{1+\Delta}$. 
The dynamical correlation function consequently obeys the following scaling form in the steady state \cite{Schmittmann:1995}:
\begin{equation} \label{eqn:KLSscalingform}
C\left( x_\parallel,\vec{x}_\perp,t \right) \sim t^{-\zeta} {\hat C}\left( \tau |\vec{x}_\perp|^{1 / \nu}, x_\parallel / |\vec{x}_\perp|^{1 + \Delta}, t / |\vec{x}_\perp|^z \right) ,
\end{equation}
with the correlation length exponent $\nu$, anisotropy exponent $\Delta$, dynamic critical exponent $z$, and $\zeta = (d + \Delta - 2 + \eta) / z$.
In addition, the non-linear coupling ratio $u / g^2$ turns out to be irrelevant in the renormalization group sense, and hence flows to zero under repeated scale transformations.
For the determination of the associated critical KLS scaling exponents, one may thus set $u \to 0$, although of course this non-linearity drives the phase separation.  
Therefore, since the remaining non-linear term $\propto g$ in the KLS Langevin equation that originates from the drive and particle exclusion only affects the longitudinal spatial sector, one has $\eta = 0$, $\nu = 1/2$, and $z = 4$, as in the corresponding Gaussian model. 
Moreover, the single remaining independent critical anisotropy exponent is fixed by Galilean invariance, as an emerging symmetry on the continuous coarse-grained level, to be $\Delta = (8 - d) / 3$, which is larger than its mean-field value $\Delta = 1$ in dimensions below the upper critical dimension $d_c = 5$.
Consequently $\zeta = (d + \Delta - 2) / 4 = (d + 1) / 6$, equal to the order parameter growth exponent $\beta = z \nu \zeta / 2 = \zeta$. 
These values of the scaling exponents for the critical KLS system were originally determined through careful field-theoretical analysis in 
Refs.~\cite{Leung:1986,Janssen:1986}; in Table~\ref{tab:table} we list them for the two-dimensional case that we consider in this work. 
The detailed procedure for how the scaling exponents and the general scaling form can be obtained from the analysis of the Langevin equation is, e.g., presented in Ref.~\cite{TauberBook:2014}.

For simpler non-critical driven diffusive systems such as the (T)ASEP, i.e., also the high-temperature phase of the KLS model, that are still governed by generic scale invariance, Eq.~\eqref{eqn:KLSscalingform} reduces to 
\begin{equation} \label{eqn:TASEPscalingform}
C\left( x_\parallel,\vec{x}_\perp,t \right) \sim t^{-\zeta} {\hat C}\left( x_\parallel / |\vec{x}_\perp|^{1 + \Delta}, t / |\vec{x}_\perp|^z \right) ,
\end{equation}
where here $\zeta = (d + \Delta + z - 2 + \eta) / z = (d + \Delta) / 2$, since of course again $\eta = 0$ and $z = 2$ as in the corresponding Gaussian approximation.
The upper critical dimension is now $d_c = 2$, and the exact generic scaling exponents become $\Delta = (2 - d) / 3$ and $\zeta = (d + 1) / 3$ in dimensions $d \leq 2$.
We note that in one dimension, Eq.~\eqref{eqn:TASEPLangevin} becomes identical to the noisy Burgers equation, and via the identification $s = - \nabla h$ also to the Kardar--Parisi--Zhang equation \cite{Kardar:1986, Kriecherbauer:2010} for the height fluctuation field $h$.
At $d_c = 2$, one obtains logarithmic corrections to the mean-field scaling exponents $\Delta = 0$ and $\zeta = 1$ (which are not explicitly listed in Table~\ref{tab:table}).

\begin{table}[t]
\begin{tabular}{lclclclc}
\hline
\hline 
\rule{0pt}{2.5ex}   & \hspace{0.6cm} $\Delta$ & \hspace{0.9cm}$z$ & \hspace{0.9cm}$\nu$ & \hspace{0.9cm}$\eta$ & \hspace{0.9cm}$\zeta$\hspace{0.4cm} \\
\hline \vspace{1mm}
\rule{0pt}{2.5ex}Critical KLS & \hspace{0.7cm}2 & \hspace{0.9cm}4 & \hspace{0.9cm}1/2 & \hspace{0.9cm}0 & \hspace{0.5cm}1/2 \\
\rule{0pt}{2.5ex}(T)ASEP & \hspace{0.7cm}0 &\hspace{0.9cm}2 & \hspace{0.9cm}-- & \hspace{0.9cm}0 &\hspace{0.5cm}1\\
\hline
\hline
\end{tabular}
\caption{\label{tab:table}Scaling exponents for the critical KLS and (T)ASEP models in two dimensions (omitting the logarithmic corrections for the (T)ASEP).}
\end{table}

\subsection{\label{sec:sublevel25}Physical Aging Scaling}
Following a rapid quench from an initial configuration that is quite distinct from the asymptotic stationary state, a system is said to be in a physical aging scaling regime if the following three properties hold:
slow (i.e., non-exponential) relaxation, broken time translation invariance, and dynamical scaling \cite{Pleimling:2010}. 
In the KLS model all three signatures of physical aging scaling are observed when the system is quenched to the critical point from a completely disordered state \cite{Daquila:2012}. 
In contrast, all distinct (T)ASEP microstates are equiprobable, whence the (T)ASEP physical aging scaling window is best accessed when the initial and final states differ drastically, e.g., when the simulation is initiated with an alternating ``checker board'' particle distribution \cite{Daquila:2011}. 
In the non-equilibrium relaxation regime, the two-time auto-correlation functions for both the (T)ASEP and KLS (at $T_c$) driven lattice gases then follow a simple aging scaling form:
\begin{equation}\label{eqn:simplescaling}
C(\vec{x}=0;t,t_w) = t_w^{- \zeta} \, {\hat C}(t / t_w) ,
\end{equation}
where $t_w$ denotes the ``waiting'' time \cite{Daquila:2011,Daquila:2012}. 
The dependence of the correlation function on both $t_w$ and $t$, not just on time difference $\tau = t-t_w$, signifies the breaking of the time-translation invariance.

Investigating physical aging phenomenona has proven especially useful for systems with conserved order parameter fields, since their aging scaling exponents can be related to the corresponding asymptotic steady state exponents \cite{Janssen:1989}. 
Therefore, studying physical aging scaling serves as an independent way to obtain the non-equilibrium critical exponents. 
This becomes exceptionally useful for systems that display exceedingly slow relaxation towards their stationary states, which is in fact the case for the critical KLS model. 

Measuring the auto-correlation function in Eq.~\eqref{eqn:simplescaling} in our simulations, we have confirmed that the KLS driven lattice gas is governed by the (T)ASEP scaling exponents in the disordered phase. 
Remarkably, we observe the scaling of $C(t,t_w)$ with (T)ASEP aging scaling exponents already at $T = 1.0 > T_c^{KLS}(E,L\to\infty) \approx 0.8$. 
The fact that the KLS driven lattice gas relaxes akin to a fully disordered system with (T)ASEP non-equilibrium scaling exponents already at temperatures not far above $T_c$ gave us the motivation to look for the dynamics of spatially inhomogeneous KLS systems with two interfaces governed by local temperature gradients. 


\section{\label{sec:level3}Two-Temperature KLS Model With Transverse Temperature Interface}

\subsection{\label{sec:sublevel31}Model Description}
\begin{figure}[t!]
\centering
\subfigure[]{\label{fig:torus}\includegraphics[width=60mm]{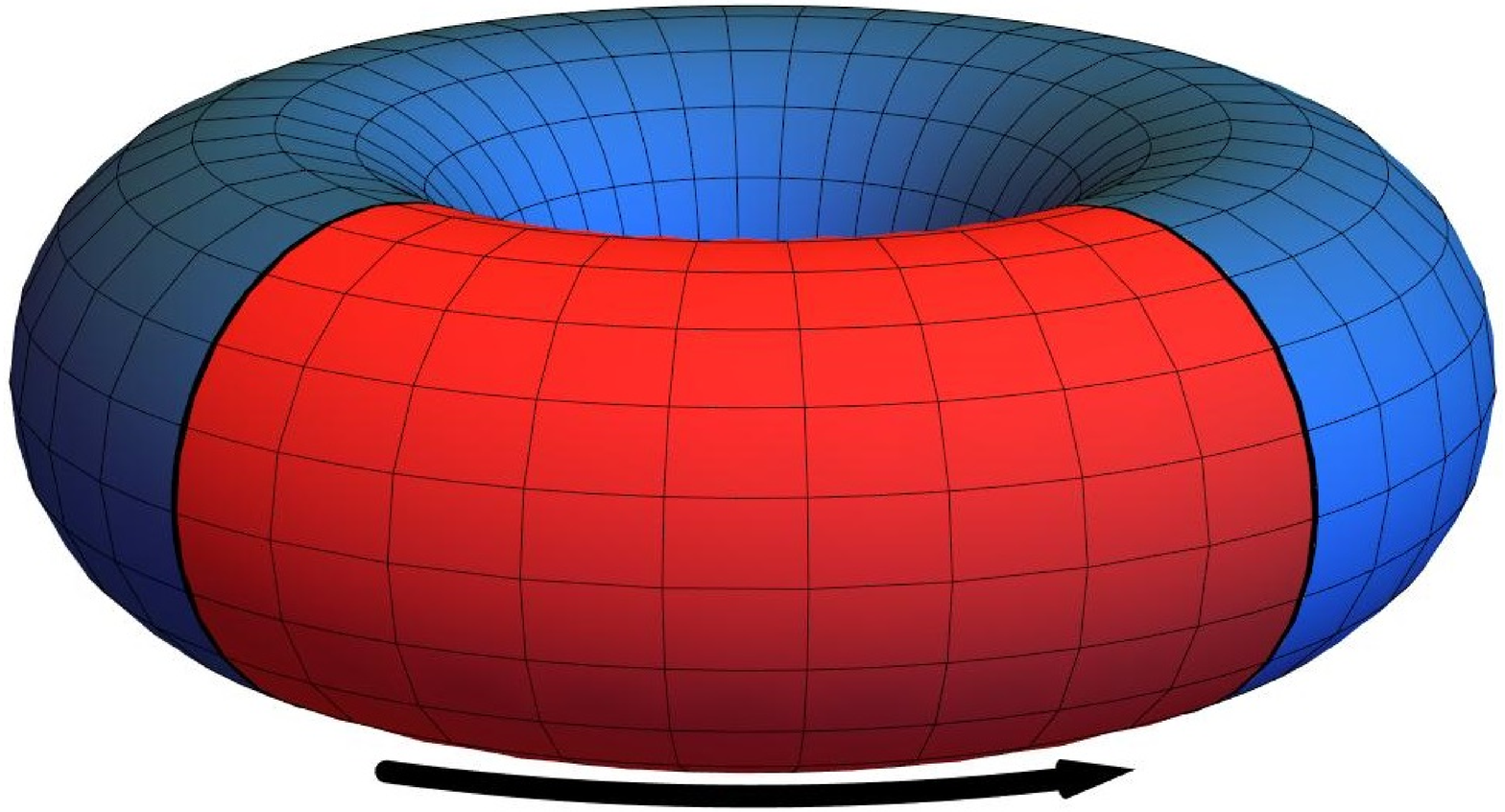}} 
\subfigure[]{\label{fig:schem}\includegraphics[width=0.9\columnwidth]{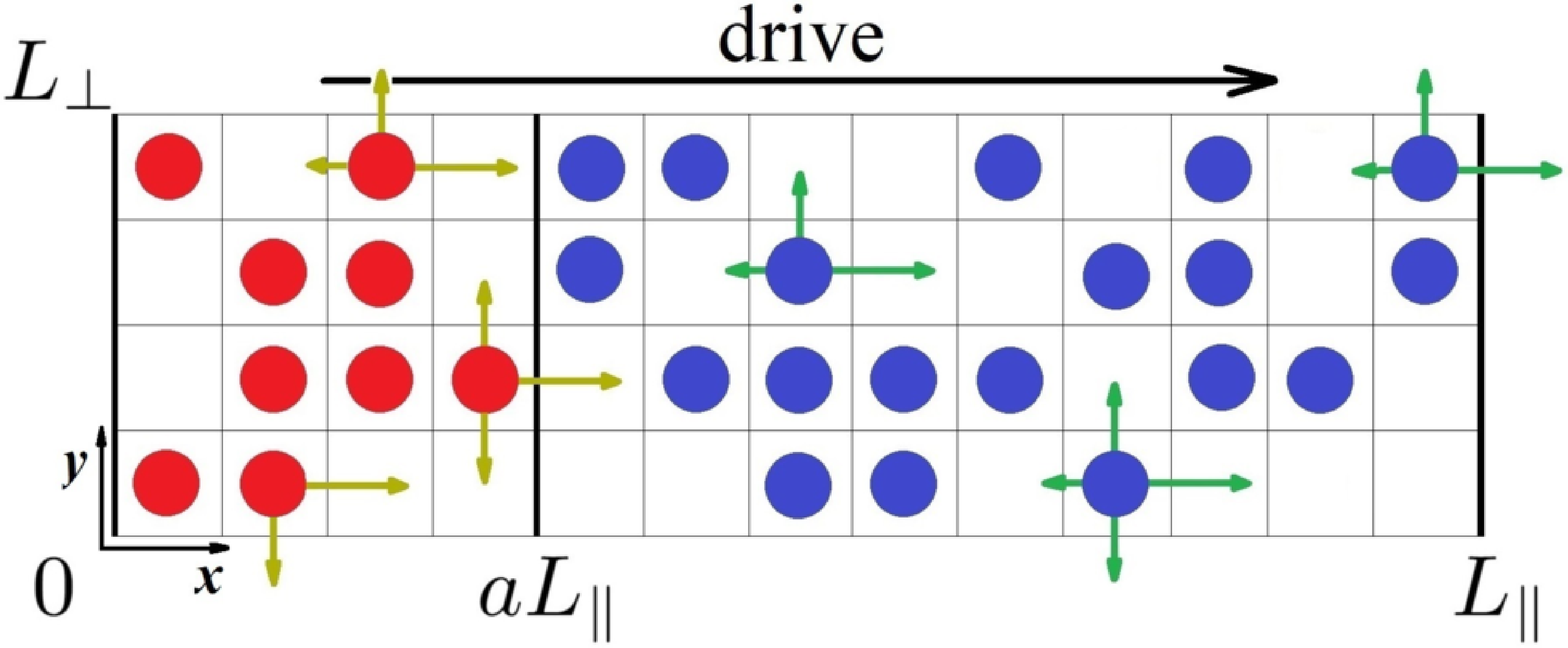}
}
\caption{\label{fig:schematics}(a) The two-temperature driven KLS lattice gas on a ring torus. 
The red sector of the torus is coupled to a reservoir at temperature $T_h > T_c$, while the blue sector is coupled to a reservoir at the critical temperature $T_l = T_c$. 
The black arrow indicates the direction of the drive. 
(b) The two-temperature driven KLS lattice gas on a two-dimensional square lattice with periodic boundary conditions. 
The $[0,aL_{\parallel})$ region of the lattice (with red-colored particles) is maintained at $T_h > T_c$, while the rest of the lattice is held at $T_l = T_c$.  
The colored arrows indicate possible hopping processes, with rates given by Eq.~\eqref{eqn:KLSrates}.}
\end{figure}

The two-temperature KLS model is a composite of two driven KLS lattice gases held at different temperatures on a ring torus, depicted schematically in Fig.~\ref{fig:torus}. 
The two lattices are coupled by allowing particle exchange across the two interfaces. 
Alternatively, the model can be cast on a $L_\parallel \times L_\perp$ rectangular square lattice with periodic boundary conditions, where the extensions of the two subsystems in the ``$\parallel$'' direction along the drive are set by the parameter $a$ as shown in Fig.~\ref{fig:schem}. 
In the first subsystem with $x \in [0, aL_{\parallel})$ with aspect ratio $0 < a < 1$, the temperature is taken to be $T_h > T_c$; whereas the second subsystem in the range $x \in [aL_\parallel, L_\parallel)$ is set, if not otherwise stated, at the critical temperature $T_l = T_c$. 
We shall henceforth refer to the two KLS subsystems according to their temperature: 
The region at $T_h > T_c$ temperature will be called ``\textit{hot}'' or alternatively the ``\textit{TASEP-like}" subsystem, and we will name the region held at $T_c$ the ``\textit{critical}'' subsystem. 
We will also refer to the single-temperature system as the ``\textit{standard}" KLS model.

In our simulations, we fix the total density to $\rho = \tfrac12$ in the two-temperature driven KLS lattice gas, in order to avoid triggering kinetic waves in the TASEP-like subsystem, and to be able to access the continuous non-equilibrium phase transition in the critical KLS subsystem. 
We also choose the drive strength to be (formally) $E = \infty$ to reduce the crossover time necessary to approach the non-equilibrium steady state. 
This $E\to\infty$ limit forbids all hops against the drive, regardless of any particle's nearest-neighbor configuration, and hence corresponds to the completely biased nearest-neighbor hopping in the longitudinal direction. 
Note that consequently the temperatures in either heat bath only affect particle movements transverse to the drive.
While the particular choice of the drive strength affects the values of the critical temperature, net stationary particle current, and overall time scale, it does not qualitatively change the KLS non-equilibrium steady state provided the drive strength satisfies $\beta E \gg 1$ \cite{Marro:1999}. 

\subsection{\label{sec:sublevel32}Monte Carlo Simulations}
We simulate the dynamics of the two-temperature driven KLS lattice gas on a two-dimensional torus or square lattice with periodic boundary conditions, see Fig.~1, using the standard Metropolis algorithm with conserved Kawasaki exchange dynamics. 
If not otherwise stated, the simulations are initiated with a random distribution of $N$ particles over the entire $L_\parallel \times L_\perp$ lattice, and proceed with random sequential updates. 
Performing $L_\parallel \times L_\perp$ updates per single Monte Carlo step (MCS), we allow every particle to be selected once on average for the update. 
If a chosen lattice site is occupied, we proceed with randomly selecting the hop direction from the four possible nearest-neighbor target sites with probability $1/4$. 
The proposed hop is then performed to an empty lattice site with a probability that depends on the drive orientation and strength, as well as on the nearest-neighbor configurations of both departure and arrival sites.
The acceptance probability for the proposed move is set by
\begin{equation}\label{eqn:HopProb}
P(\mathcal{C}\to \mathcal{C'}) = \min\left\{ 1, \exp\left( -\beta [H(\mathcal{C'}) - H(\mathcal{C}) - l E] \right) \right\}  ,
\end{equation}
where $H(\mathcal{C})$ and $H(\mathcal{C'})$ are the energy of the initial and final configurations, respectively, computed from Eq.~\eqref{eqn:IsingHamilt}.

We use a different expression for the acceptance probability for the hops across the temperature boundaries:
\begin{align}\label{eqn:HopProb_bc}
\begin{split}
P&(\mathcal{C}\to \mathcal{C'})\big|_{\beta_1 \to \beta_2} = \\
=& \min\left\{ 1, \exp\left( -\beta_2 H(\mathcal{C'}) + \beta_1 [H(\mathcal{C}) + l E] \right) \right\} ,
\end{split}
\end{align} 
where $\beta_1$ and $\beta_2$ denote the inverse temperatures pertinent to the particle's initial and final position. 
In the $E \to \infty$ limit, simply all hops along the drive will be accepted, while all hops against the drive are strictly prohibited; the Ising Hamiltonian consequently only affects hops transverse to the drive direction.
Since the temperature interfaces in this work are oriented perpendicular to the drive, our particular choice \eqref{eqn:HopProb_bc} to handle the hops across the subsystem boundaries in fact would not matter.
We will henceforth measure the temperature $T$ that solely controls the probability of particle motion transverse to the drive in units of $\mathcal{J} / k_{\rm B}$.

\section{\label{sec:level4}Transient regime}

\begin{figure}[t!]
\centering
\vspace{2 mm}
\subfigure[$\, 3,000$ MCS]{\includegraphics[width=0.985\columnwidth]{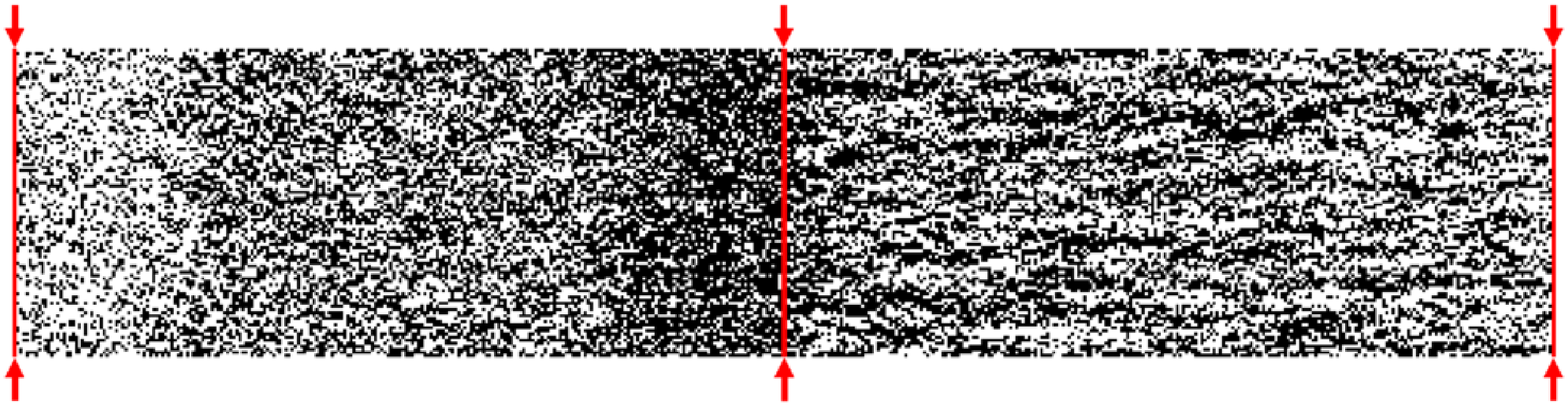}\label{fig:snapshot_3k}
}\\[0.5ex]

\subfigure[$\, 10,000$ MCS]{\includegraphics[width=0.985\columnwidth]{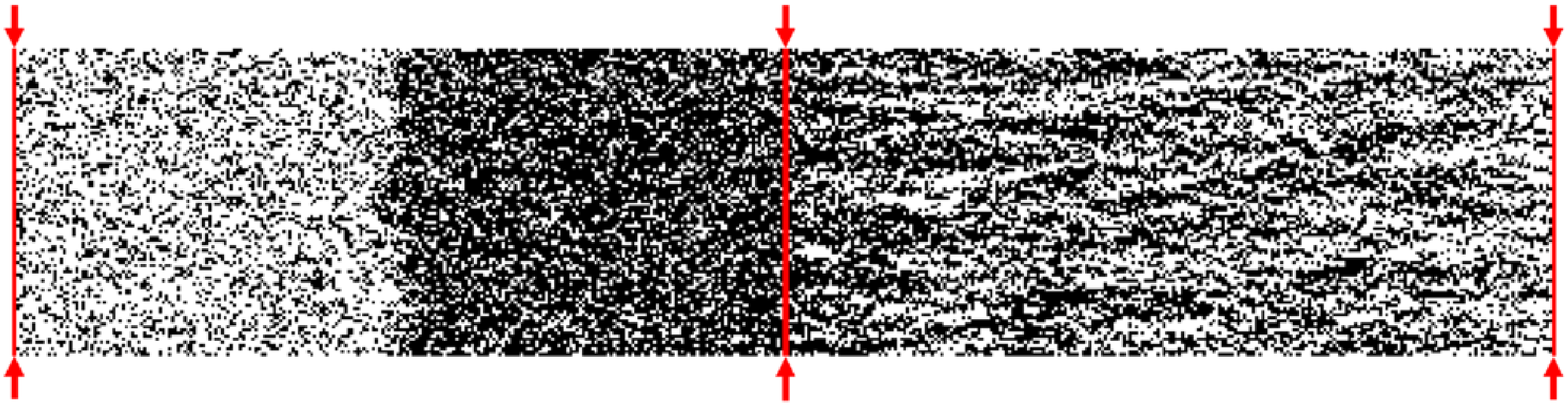}\label{fig:snapshot_10k}}
\caption{\label{fig:sine}Simulation snapshots of the two-temperature KLS driven lattice gas with $L_\parallel =  500$ and $L_\perp = 100$ after 
(a) $3,000$ Monte Carlo steps (MCS) and (b) $10,000$ MCS. 
The vertical red lines and arrows indicate the location of the two temperature interfaces.
The left part of the lattice in the figure is coupled to the temperature bath at $T_h = 2.0$, and the right part to the $T_l = 0.8 \approx T_c$ reservoir.}
\end{figure}
In order to access the physical aging scaling regime in the hot or TASEP-like subsystem (held at temperature $T_h > T_c)$, we start from highly correlated initial conditions and fill the whole lattice with particles in an alternating manner, i.e., a checker-board pattern. 
Shortly after the beginning of the simulation, we observe the formation of two density shock waves in the hot subsystem, as seen in the simulation snapshot in Fig.~\ref{fig:snapshot_3k}: 
The low-density shock wave nucleates at the boundary (at $x = 0$ and $x = L_\parallel$) that particles cross to enter the hot subsystem from the cooler side where $T_l = T_c$; the high-density shock wave emerges at the interface (located at $x = aL_\parallel$) where particles leave the TASEP-like subsystem. 
These high- and low-density shocks traverse the hot region, moving toward each other with the same constant velocity; its value is consistent with the one determined for the one-dimensional TASEP with open boundaries in the coexistence phase \cite{Popkov:1999}: 
\begin{equation}
v = \frac{\langle J_{\parallel, st}\rangle - \langle J_{\parallel, st}(T_h)\rangle}{ \rho_\pm - \rho} \, ,
\label{eqn:vel}
\end{equation}
where $\langle J_{\parallel, st}(T_h)\rangle$ is the mean drive-induced steady-state particle current in the \textit{standard} KLS model at temperature $T_h$, whereas $\langle J_{\parallel, st}\rangle$ represents the mean drive-induced steady-state particle current in the \textit{two-temperature} KLS system; $\rho_\pm$ denote the average high and low densities on either side in the hot subsystem, and $\rho = \tfrac12$ is the total density in the lattice. 

\begin{figure}[t!]
\centering
\includegraphics[width=\columnwidth, trim={0.3cm 0 1.5cm 2cm},clip]{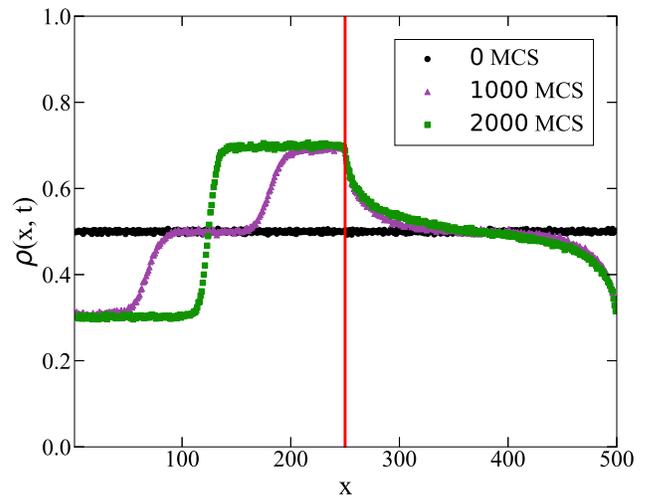}\\[-2.4ex]
\caption{\label{fig:DenProf}Density profiles of the two-temperature KLS driven lattice gas with dimensions $L_\parallel = 500$, $L_\perp = 250$ at different simulation times. 
The subsystem length ratio is 1:1 ($a = 0.5$); one temperature interface is indicated by the vertical red line, while the other one is located at $x = 0$ and $x = L_\parallel$. 
The hot subsystem is held at $T_h = 2.0$ and the critical subsystem at $T_l = 0.8 \approx T_c$. 
The data are averaged over $100$ independent realizations.}
\end{figure}
Once the two density shock waves have met in the middle of the hot subsystem, a stationary domain wall forms that completely separates the subsystem into low- and high-density phases, as shown in the late-time snapshot, Fig.~\ref{fig:snapshot_10k}. 
Recording the density profile for the two-temperature KLS model enables us to follow the temporal evolution of the system, and allows us to obtain the kink's height and width as shown in Fig.~\ref{fig:DenProf}. 
We will discuss the intriguing steady-state shape of the density profile in the following section~\ref{sec:level5}.
Meanwhile, the critical subsystem (at temperature $T_ l = 0.8 \approx T_c$) develops long correlated clusters that are oriented along the drive. 
In contrast to the standard KLS model, the stripe-like clusters in the critical region of the two-temperature KLS system become more dense at the boundary where particles enter the critical domain, and less dense at the boundary where particles leave this region. 
The extended correlated particle clusters continue to alter their shape even after the two density shock waves have met one another in the hot region, eventually forming funnel-like structures that traverse the entire critical subsystem. 
The process of the long density tails spreading into the critical subsystem happens on a much longer time scale, which we have not been able to reliably estimate; perhaps this kinetics is governed by a power law. 

\begin{figure}[t!]
\centering
\includegraphics[width=\columnwidth, trim={0 0 1.5cm 2cm},clip]{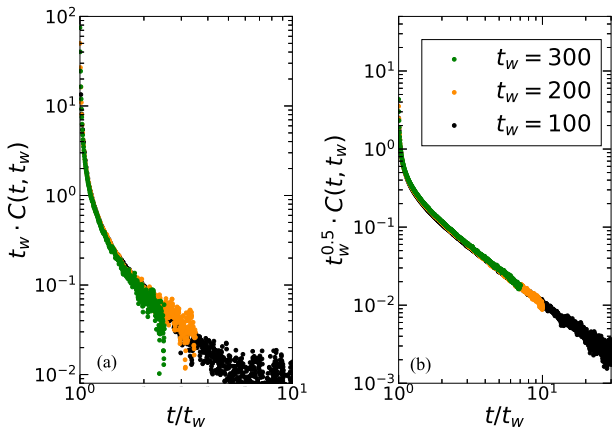}\\[-2ex]
\caption{\label{fig:Aging}Aging scaling plots of the two-time density auto-correlation function in the two-temperature KLS driven lattice gas with dimensions $L_\parallel =  1000$, $L_\perp = 64$. 
The data are collected following a quench from a random initial state to the high-temperature state ($T_h = 5.0$) in the left graph, and to the critical point ($T_l = 0.8 \approx T_c$) in the right plot, in the central columns of the hot and critical subsystems, respectively. 
The points in each curve represent averages over $1,000,000$ independent realizations.}
\end{figure}

The emergence of these spatial inhomogeneities in the two-temperature KLS driven lattice gas prevents us from accessing the physical aging scaling regime in the parts of the lattice with non-uniform density. 
However, the density waves nucleate at the temperature interface boundaries and the approximate time it takes for them to reach the center of the hot subsystem is $\Delta t \approx aL_\parallel / 2v$, since $aL_\parallel$ is the length of the hot subsystem along the drive direction.
Until that moment, the central parts of the two subsystems do not experience any mutual coupling, and are effectively independent.
One would hence expect the initial non-equilibrium relaxation dynamics in the TASEP-like subsystem to be governed by the TASEP aging scaling exponents, and  the critical subsystem correspondingly characterized by the critical KLS aging exponents. 
To verify this assertion, we have probed the aging scaling exponent $\zeta$ separately in the middle columns of both the hot and critical subsystems. 
Applying the simple aging scaling form \eqref{eqn:simplescaling}, we demonstrate in Fig.~\ref{fig:Aging} that the obtained aging scaling exponents from the central part of the hot and critical subsystems are indeed identical to the predicted and reported values for the two-dimensional (T)ASEP ($\zeta = 1$) and standard KLS ($\zeta = 1/2$) models, respectively \cite{Daquila:2011, Daquila:2012}. 

\begin{figure*}[t!]
\centering
\subfigure[\label{fig:J_T}]{\includegraphics[width=\columnwidth, trim={0 0.5cm 1.5cm 2cm},clip]{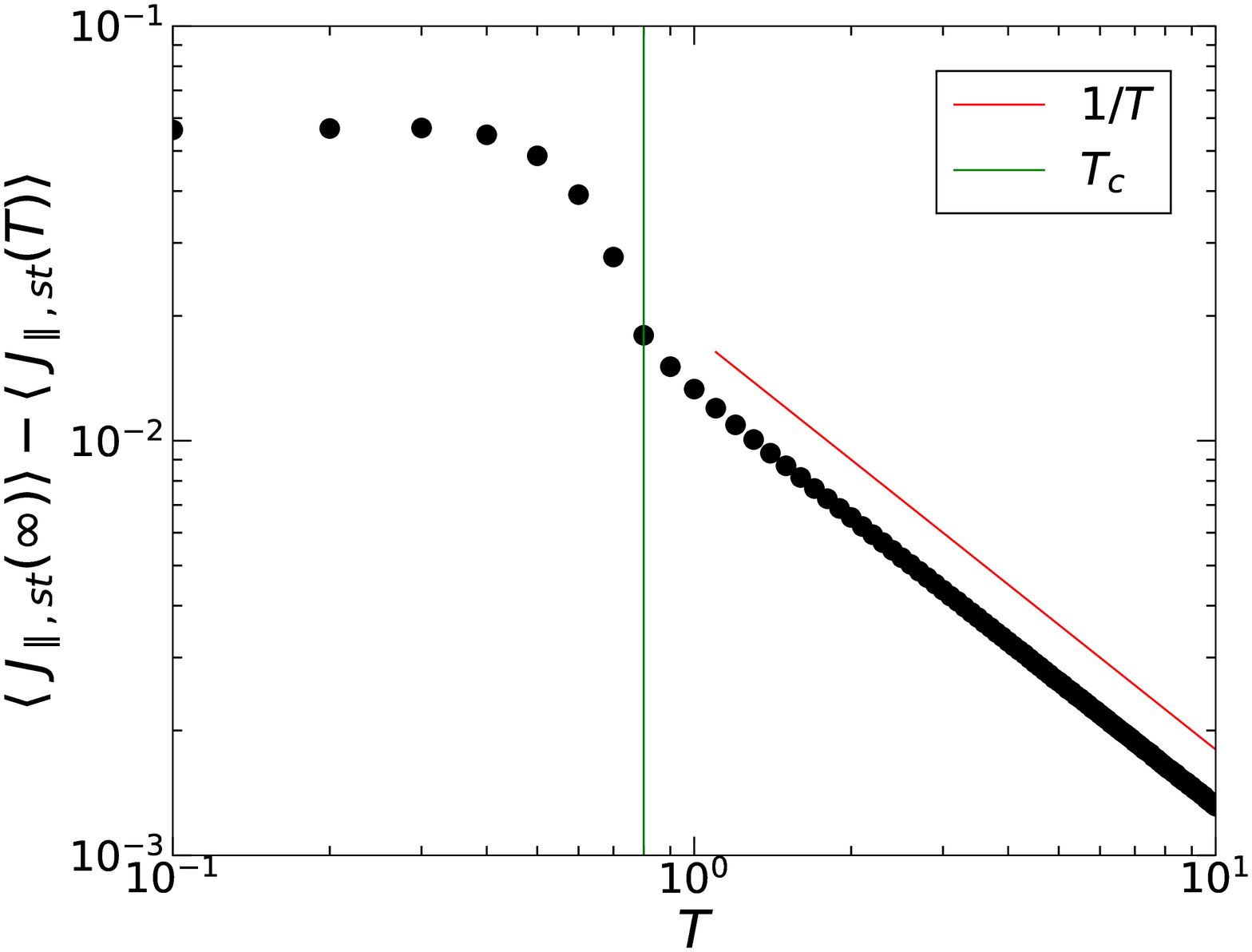}}
\hfill
\subfigure[\label{fig:CurProf}]{\includegraphics[width=\columnwidth, trim={0 0.5cm 1.5cm 2cm},clip]{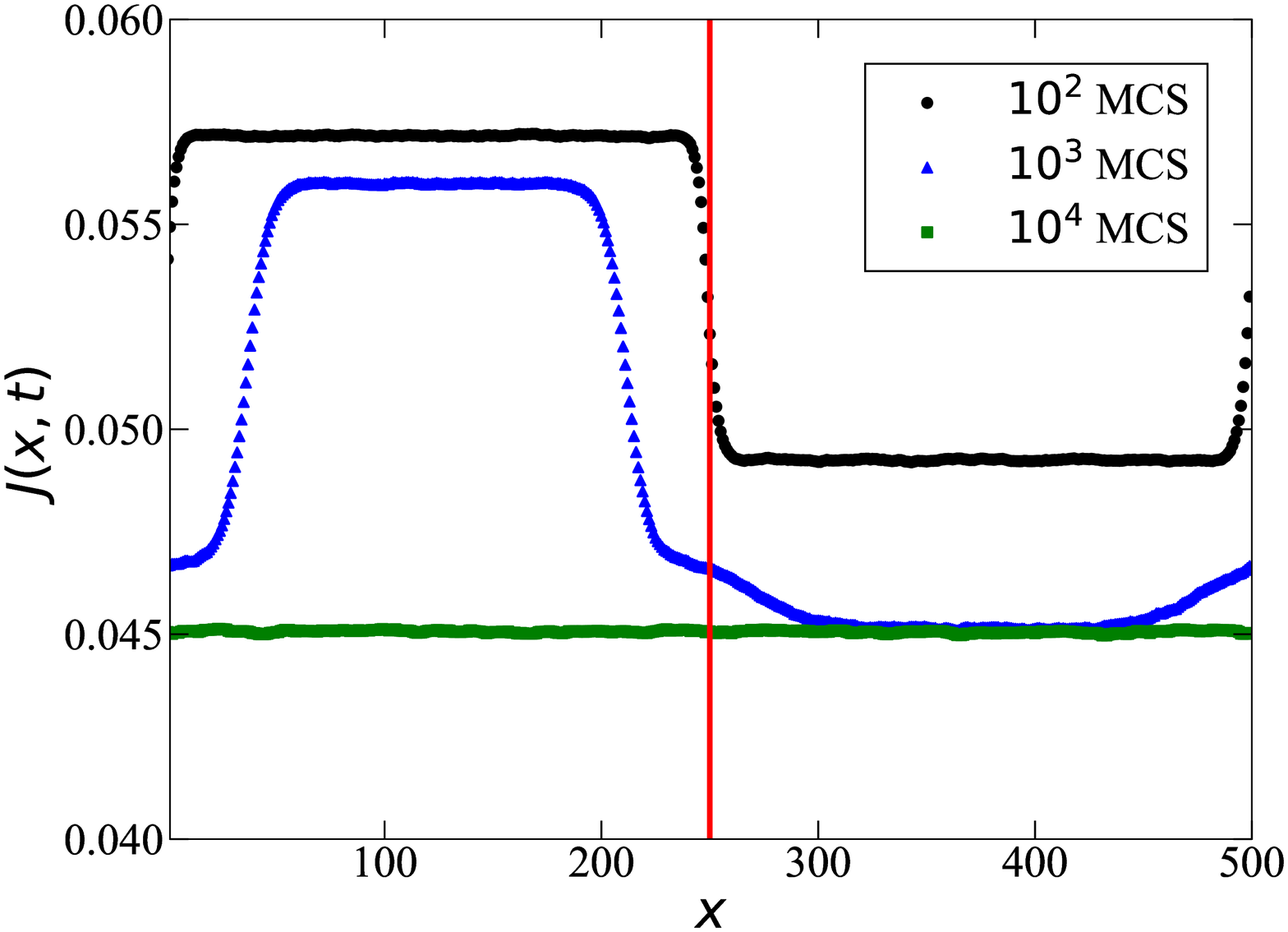}}
\caption{(a) The drive-induced current at different temperatures in the standard (uniform) KLS driven lattice gas with dimensions $L_\parallel = 1000$, $L_\perp = 64$. 
The data are collected after $20,000$ MCS when the system is in the steady state for $T > T_c$, but still in a very slowly evolving transient state for $T \leq T_c$. 
The data are averaged over $1000$ MCS and over $100$ independent realizations. 
(b) Transient current profiles of the two-temperature KLS driven lattice gas with dimensions $L_\parallel =  500$, $L_\perp = 250$ at different times. 
The subsystem ratio is 1:1 ($a = 0.5$) and one of the temperature interfaces is indicated by the red line. 
The hot subsystem is held at $T_h = 2.0$ and the critical subsystem at $T_l = 0.8 \approx T_c$. 
The data are averaged over $100$ MCS and over $1,000$ independent realizations.}
\end{figure*}
Once the two density shock waves meet at the center of the TASEP-like subsystem, the initial aging scaling terminates, and the subsystem becomes completely segregated into the high- and low-density regions, with the single domain wall centered in the middle of the hot subsystem. 
This phase separation in the high-temperature region can be explained by considering the mean drive-induced particle current in the two-temperature KLS system. 
However, before examining this inhomogeneous two-temperature modification of the KLS model, it is useful to first analyze how the mean steady-state particle current depends on the temperature in the original standard KLS driven lattice gas. 
To our knowledge, only Katz et al. have studied this temperature dependence of the mean steady-state particle current in their original paper \cite{Katz:1984}. They showed that the mean steady-state current $\langle J_{\parallel, st}(T) \rangle$ assumes the value $p_\parallel \rho (1-\rho)$ in the infinite-temperature limit, and drops drastically in the ordered phase due to the stripe-like phase separation of the system, with a cusp discontinuity at the critical temperature $T_c$. 
Seeking to uncover the functional dependence of $\langle J_{\parallel, st}(T) \rangle$ on $T$, we found that the mean steady-state current in the standard KLS model decays linearly with inverse temperature $\langle J_{\parallel, st}(\infty) \rangle - \langle J_{\parallel, st}(T) \rangle \sim T^{-1}$ in the disordered phase ($T > T_c$), as shown in Fig.~\ref{fig:J_T}. 
This inverse temperature power law clearly originates from the decrease of the effective Ising attractive interactions with $T$.

Since the two-temperature KLS driven lattice gas must of course reduce to the standard KLS model for $a = 0$ or $1$, one would expect the mean steady-state particle current $\langle J_{\parallel, st} \rangle$ in the two-temperature KLS system at temperatures $T_h$ and $T_l < T_h$ to be bounded between the standard KLS values of the mean steady-state current $\langle J_{\parallel, st}(T_h) \rangle$ and $\langle J_{\parallel, st}(T_l) \rangle$. 
We have confirmed this expectation with our simulations, and show in Fig.~\ref{fig:CurProf} that initially the hot and critical subsystems have distinct local current values.
But once the two-temperature KLS driven lattice gas has reached its steady state, the particle current $\langle J_{\parallel, st} \rangle$ becomes uniform across the system and takes values in the range $[ \langle J_{\parallel, st}(T_l) \rangle , \langle J_{\parallel, st}(T_h) \rangle ]$. 
One may draw an analogy between the two-temperature KLS driven lattice gas and the TASEP with a slow bond to illustrate how the drop in the current causes a particle blockage in the two-temperature KLS model. 
In the TASEP with a slow bond, the defect bond plays the role of a bottleneck that impedes the transport in the entire chain. 
Similarly, in the two-temperature KLS driven lattice gas the entire critical subsystem serves as the bottleneck for the particles emanating from the hot region. Once these particles enter the critical subsystem, they become stuck inside the large correlated clusters. 
This causes the clogging at the hot-to-critical temperature interface, and eventually induces the density phase separation in the hot subsystem. 

Indeed, we observe the phase separation inside the hot subsystem even when the temperature $T_l$ of the cooler subsystem is above the critical temperature, $T_h > T_l > T_c$: 
When both subsystems are in the disordered phase, the subsystem with the lower temperature will play the role of the bottleneck region since according to Fig.~\ref{fig:J_T} it sustains a lower stationary mean particle current.


\section{\label{sec:level5}Steady-State Properties}
\subsection{\label{sec:sublevel51}Steady-State Particle Current}
Once the density profile ceases altering its shape and the drive-induced particle current becomes uniform across the whole lattice and does not change with time anymore, the two-temperature KLS driven lattice gas has reached its non-equilibrium stationary state.
The mean steady-state particle current $\langle J_{\parallel, st}\rangle$ in the two-temperature KLS driven lattice gas depends non-trivially on all system parameters: the system's geometry determined by $L_\parallel$, $L_\perp$ and the aspect ratio $a$, as well as both temperatures $T_h > T_l$. 
As we have mentioned in the preceding section, these temperatures of the hot and cooler subsystems $T_l$, $T_h$ set the upper and lower boundaries for $\langle J_{\parallel, st} \rangle$ that are approached when the parameter $a$ either takes the value $1$ or $0$ (rendering the system uniform).

\begin{figure}[b!]
\centering
\includegraphics[width=\columnwidth, trim={0 0.5cm 0.5cm 2cm}]{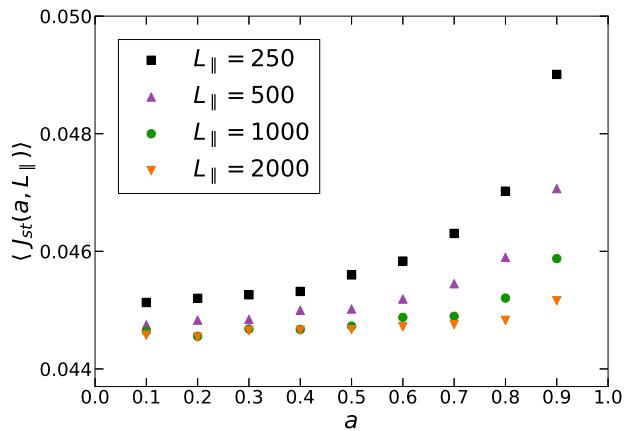}
\caption{The steady-state current for different subsystem ratios and overall system sizes. 
The aspect ratio parameter $a$ indicates the position of the hot-to-critical subsystems' interface. 
The hot subsystem was held at $T_h = 2.0$, while the critical subsystem was maintained at $T_l = 0.8 \approx T_c$. 
The lattice width $L_\perp=64$ was used for the simulations, and the data were averaged over $5,000$ MCS and over $10$ independent realizations.}
\label{fig:SteadCur}
\end{figure}

We observe the two-temperature KLS model to display intriguing and subtle finite-size features. 
For example, in Fig.~\ref{fig:SteadCur} we show how the mean steady-state particle current $\langle J_{\parallel, st} \rangle$ varies with the aspect ratio $a$ for different total lattice lengths $L_\parallel$. 
As in the TASEP with a slow bond, $\langle J_{\parallel, st} \rangle$ drops significantly even when even just a small fraction $1 - a \ll 1$ of the lattice is maintained at the critical temperature, or in fact at any $T_l$ with $T_c < T_l < T_h$; i.e., the lattice sites that are coupled to the lower-temperature bath dominantly affect the resulting mean stationary current values. 
For the same subsystem ratio $a$ but different $L_\parallel$, the mean steady-state particle current is be lower in systems with a greater number of effectively slow lattice sites controlled by the heat bath set to temperature $T_l$; hence $\langle J_{\parallel, st} \rangle$ becomes reduced for larger $L_\parallel$.

In contrast to the TASEP with a slow bond, the ``defect'' region in the two-temperature KLS driven lattice gas emerges in the critical subsystem owing to the Ising attractive interactions that instigate the formation of the long horizontal clusters. 
To gain better insight into the steady-state particle current dependence on the system size $L_\parallel$, we have analyzed single-row wide clusters in the critical subsystem.
We apply the same rules to define a cluster that Katz et al. used in their original work \cite{Katz:1984} to study the standard KLS clusters size histogram: 
A single-row wide particle cluster is assigned a length $n$ if it has empty sites at its ends, and no empty sites inside the cluster. 
Thus, according to this definition, a single separate particle is considered to be a cluster of size $n=1$, and a completely occupied row corresponds to a cluster of size $n = L_\parallel$. 

\begin{figure}[t!]
\centering
\includegraphics[width=\columnwidth, trim={0.cm 0.cm 0cm 0cm}]{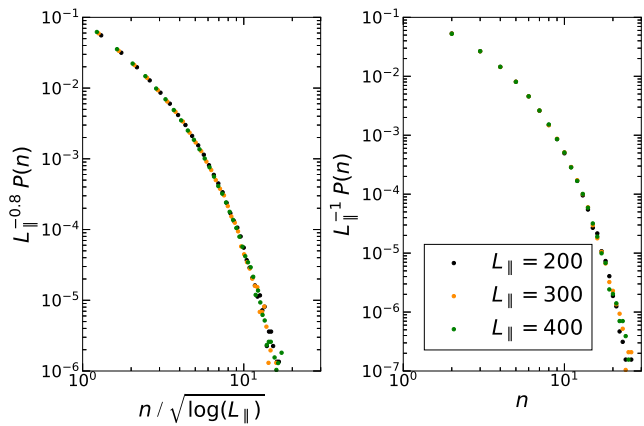}
\caption{Double-logarithmic plots of the probability distribution $P(n)$ of the cluster sizes $n$ in the cooler subsystem held at (a) $T_l = 0.8 \approx T_c$, and (b) $T_l = 2.0 > T_c$, while the hot region is maintained at (a) $T_h = 2.0$, and (b) $T_h = 5.0$, respectively. 
The lattice width $L_\perp = 32$ was used for the simulations. 
The data for each curve were collected in the steady state after $L_\parallel^2$ MCS and were averaged over $1,000$ independent realizations.}
\label{fig:ClusterStat}
\end{figure}

After reaching the steady state in our simulation, we have collected all clusters in the critical subsystem, which is held at $T_c$, into $n$-sized bins, and subsequently constructed their histogram.
Thereby arriving at an estimate for the probability distribution of the cluster size $P(n)$, we have found the following scaling with system size $L_\parallel$:
\begin{equation}
P(n) = L_\parallel^\alpha \, \mathcal{F}\left( n  / \xi_\parallel(L_\parallel) \right) ,
\label{eq:clsize}
\end{equation}
where we obtain $\alpha \approx 0.8$ from the optimal data collapse, and the characteristic correlation length or typical cluster size $\xi_\parallel(L_\parallel) \sim \sqrt{\log L_\parallel}$ grows logarithmically with system size $L_\parallel$, as shown in Fig.~\ref{fig:ClusterStat}(a). 
Yet if $T_l$ in the cooler subsystem is set above the critical temperature, $T_h > T_l > T_c$, the finite-size scaling exponent in Eq.~\eqref{eq:clsize} changes its value to $\alpha \approx 1$, while the correlation length $\xi_{\parallel}$ becomes constant and of $\mathcal{O}(1)$, see Fig.~\ref{fig:ClusterStat}(b). 
When the cooler subsystem resides in the KLS disordered phase, long-range spatial correlations disappear, and particles move essentially freely and unimpeded through the lower-temperature region.

From these observations we conclude that the overall decrease of the mean steady-state particle current in the two-temperature KLS model with $L_\parallel$ occurs because the characteristic cluster size in the critical subsystem increases logarithmically with the system length $L_\parallel \sim L_\perp^{1 + \Delta} = L_\perp^3$ and hence also with its width $L_\perp$. 
Moreover, the formation of these correlated critical clusters markedly affects the dynamics of the phase interface fluctuations in the hot subsystem, as we will discuss in the final subsection~\ref{sec:sublevel53} below.

\subsection{\label{sec:sublevel52}Steady-State Density Profile}

For the two-temperature KLS model, the shape of the density profile in Fig.~\ref{fig:DenProf} resembles a hyperbolic $\tanh$-function in the hot region and a trigonometric $\tan$-function in the low-temperature subsystem. 
Remarkably, hyperbolic tangent- and tangent-shaped density profiles have been observed in the one-dimensional TASEP with open boundaries in the coexistence and the maximal-current phases, respectively \cite{Derrida:1993b,Schutz:1993,Krug:1991,Schutz:2001,Blythe:2007,Stinchcombe:2011}. 
This striking similarity calls for a quantitative comparison of the two-temperature KLS steady-state density profiles with the known mean-field solutions for the one-dimensional TASEP with open boundaries, as given by Eqs.~\eqref{eqn:TASEPtanh} and \eqref{eqn:TASEPtan}.
In the following, we discuss the unique properties of the density profiles in both regions of the two-temperature KLS system. 
In particular, we wish to ascertain whether the mean-field description for the TASEP with open boundaries may be adapted to characterize the two-temperature KLS density profiles for two separate cases, namely either when the cooler subsystem is maintained at the critical temperature ($T_l = T_c$), or when it is held at a temperature just above the critical one.

\subsubsection{\label{sec:sublevel521}Critical Subsystem}

\begin{figure}[t!]
\centering
\includegraphics[width=\columnwidth, trim={0 0.5cm 1.5cm 1.5cm}]{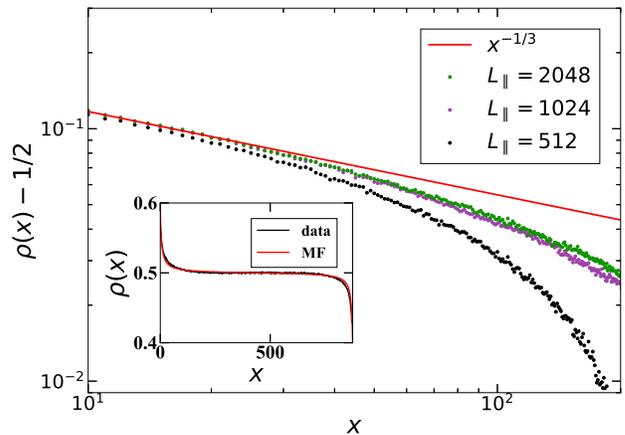}
\caption{\label{fig:KLStails}Decay of the density $\rho(x)$ in the critical subsystem, where $x$ is the distance from the hot-to-critical temperature interface. 
The hot subsystem is held at $T_h = 2.0$; the hot and critical subsystem size ratio is chosen to be 1:8. 
Different curves correspond to the different system lengths $L_\parallel$, but with identical width $L_\perp = 64$. 
The data points in each curve reflect averages over $10,000$ independent realizations. 
The inset compares the $\tan$-like part of the density profile with the mean-field result when both subsystems are maintained at temperatures above the critical one, namely at $T_h = 10.0$ and $T_l = 5.0$, respectively. The system length is $L_\parallel = 1,000$, the system width $L_\perp = 64$, and the hot and critical subsystem size ratio is chosen to be 1:8.
The data shown in the inset were averaged over $10,000$ independent realizations.}
\end{figure}
We first carefully examine the critical subsystem, which is maintained at $T_l = 0.8 \approx T_c$.
We have found that the density field $\rho(x)$ does not follow the mean-field tangent solution given by Eq.~\eqref{eqn:TASEPtan} as proposed for the maximal current phase in the TASEP with open boundary conditions. 
Instead, the excess density $\rho(x) - \langle \rho \rangle$ with $\langle \rho \rangle = \tfrac12$, near the boundaries of the critical subsystem displays a power law decay away from the interface with exponent $1/3$, as shown in Fig.~\ref{fig:KLStails}. 
This decay exponent is indeed characteristic of the critical KLS universality class, as we demonstrate with the general KLS scaling form \eqref{eqn:KLSscalingform} of the correlation function at the critical point ($\tau = 0$) and $x_\perp = 0$,
\begin{equation}
C(t,x_\parallel) = t^{-\zeta} \, \widetilde{C}(t / x_\parallel) \sim x_\parallel^{-\zeta \, z / (1 + \Delta)} 
\end{equation}
at $t = 0$.
Upon identifying $x = x_\parallel$, which now indicates the distance measured parallel to the drive direction from the hot-to-critical temperature interface (i.e., from the red line to the right in Fig.~\ref{fig:DenProf}), we thus find for the excess density 
\begin{equation}
    \big| \rho(x) - \tfrac12 \big| \sim x^{-\zeta \, z / [2 (1 + \Delta)]} = x^{-1/3}
\end{equation}
in two dimensions, with $\Delta = 2$ and $z = 4$ representing the critical KLS anisotropy and dynamical exponents, and $\zeta = \Delta / 4 = 1/2$; see Table~\ref{tab:table}. 
From the graphs pertaining to different $L_\parallel$ in the main plot on Fig.~\ref{fig:KLStails}, we infer that the deviations from this power law are due to the finite size of the system, and posit that the excess density will follow the $x^{-1/3}$ algebraic decay in the thermodynamic limit.

Our simulation data in the inset of Fig.~\ref{fig:KLStails} indicate that the density profile in the cooler subsystem follows the mean-field solution when the temperature $T_l > T_c$ in that region is raised above the critical value. 
Thus we may directly adapt Eq.~\eqref{eqn:TASEPtan} for the one-dimensional TASEP in the maximal current phase to the two-temperature KLS model, setting $x_0 = (1+a) L_\parallel / 2$.
In contrast to the TASEP formula the inverse characteristic length now is $q = \sqrt{\langle J_{\parallel, st} \rangle / J_{\rm max} - 1}$, with the mean steady-state current $\langle J_{\parallel, st}\rangle$ in the two-temperature KLS model, and where  $J_{\rm max}$ is the cooler subsystem maximal current that is equal to the mean steady-state current $\langle J_{\parallel, st}(T_l) \rangle$ in the standard KSL model at temperature $T_l$. 

\subsubsection{\label{sec:sublevel522}Hot Subsystem}

\begin{figure}[t!]
\centering
\includegraphics[width=\columnwidth, trim={0 0cm 0cm 1.5cm}]{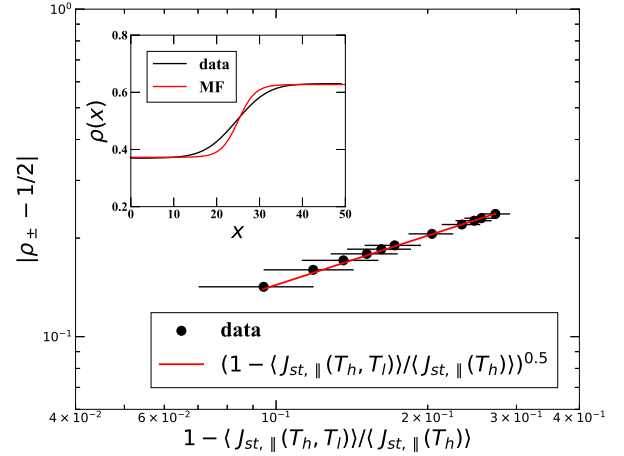}
\caption{\label{fig:rates} Double-logarithmic plot of the hot subsystem's low and high densities $\rho_\pm$ as a function of the mean steady-state particle current and the maximal current for different temperatures of the hot subsystem. 
The cooler subsystem is held at the critical temperature, $T_l = 0.8 \approx T_c$. 
The system size is $L_\parallel = 1,000$ and $L_\perp = 64$ and the hot-to-critical subsystem ratio is 1:8.
All data points were averaged over $10,000$ independent realizations. 
The inset compares the $\tanh$-like part of the density profile with the mean-field result when both subsystems are maintained at temperatures above the critical one, namely at $T_h = 5.0$ and $T_l = 2.0$, respectively; here the hot-to-critical subsystem ratio was set to 1:19 ($a = 0.05$).
The data in the inset were averaged over $10,000$ MCS and over $1,000$ independent realizations.}
\end{figure}

Our simulations show that the mean-field solution \eqref{eqn:TASEPtanh} for the one-dimensional TASEP with open boundaries in the coexistence phase can be readily adapted to the two-temperature KLS driven lattice gas to describe the density profile of the hot subsystem, with $x_ 0 = a L_\parallel / 2$.
Here the inverse characteristic length is $k = \sqrt{1 - \langle J_{\parallel, st} \rangle / J_{\rm max}}$, with the mean steady-state current $\langle J_{\parallel, st} \rangle$ in the two-temperature KLS model, and $J_{\rm max}$ denoting the hot subsystem maximal current, i.e., the mean steady-state current $\langle J_{\parallel, st}(T_h) \rangle$ in the standard KLS model at temperature $T_h$. 
This mean-field expression works exceptionally well far away from the phase interface, and if $T_h > T_l > T_c$. 
Having obtained the density profiles and the mean steady-state particle currents from our simulations for systems at different temperatures $T_h$, we show in Fig.~\ref{fig:rates} that the asymptotic high and low densities at the boundaries of the hot subsystem are completely determined by the value of the inverse characteristic length $k$,
\begin{equation}\label{eqn:asymptotes}
\langle \rho_\pm \rangle = \tfrac{1}{2}(1 \pm k) = \tfrac{1}{2} \left[ 1 \pm \left( 1 - \langle J_{\parallel, st } \rangle / J_{\rm max} \right)^{1/2} \right] .
\end{equation}
However, small but clearly noticeable deviations from this expression are observed, when both subsystems are maintained at temperatures close to $T_c$, $2.0 > T_h, T_l \gtrapprox T_c \approx 0.8$ (on the left side of the figure), signifying the influence of critical fluctuations across the entire system that are not captured by the mean-field approximation. 
Indeed, instead of the exponent $1/2$ in Eq.~\eqref{eqn:asymptotes}, our best data fit gives the value $0.48$, yet with error bars that are still compatible with the mean-field exponent $0.5$.
We also observe that the prefactor in front of the brackets in Eq.~\eqref{eqn:asymptotes} is always smaller than $\tfrac12$ in systems with $T_l \approx T_c$; i.e., the actual height of the kink in the hot subsystem is reduced as compared to the mean-field prediction.

Close to the phase interface, the domain wall width differs significantly from the mean-field prediction. 
This discrepancy clearly arises from the presence of non-trivial phase interface fluctuations that are not accounted for by the mean-field approximation. 
Examining a series of successive simulation snapshots, we have confirmed that the phase interface shape in the hot subsystem is in fact changing with time. 
The following subsection is devoted to a more detailed exploration of the interface fluctuations at the boundary between the high- and low-density phases within the high-temperature region.

\begin{figure*}[t!]
\centering
\subfigure[\label{fig:int_fluct_spectrum}]{\includegraphics[width=\columnwidth, trim={0cm 0cm 0cm 0cm}]{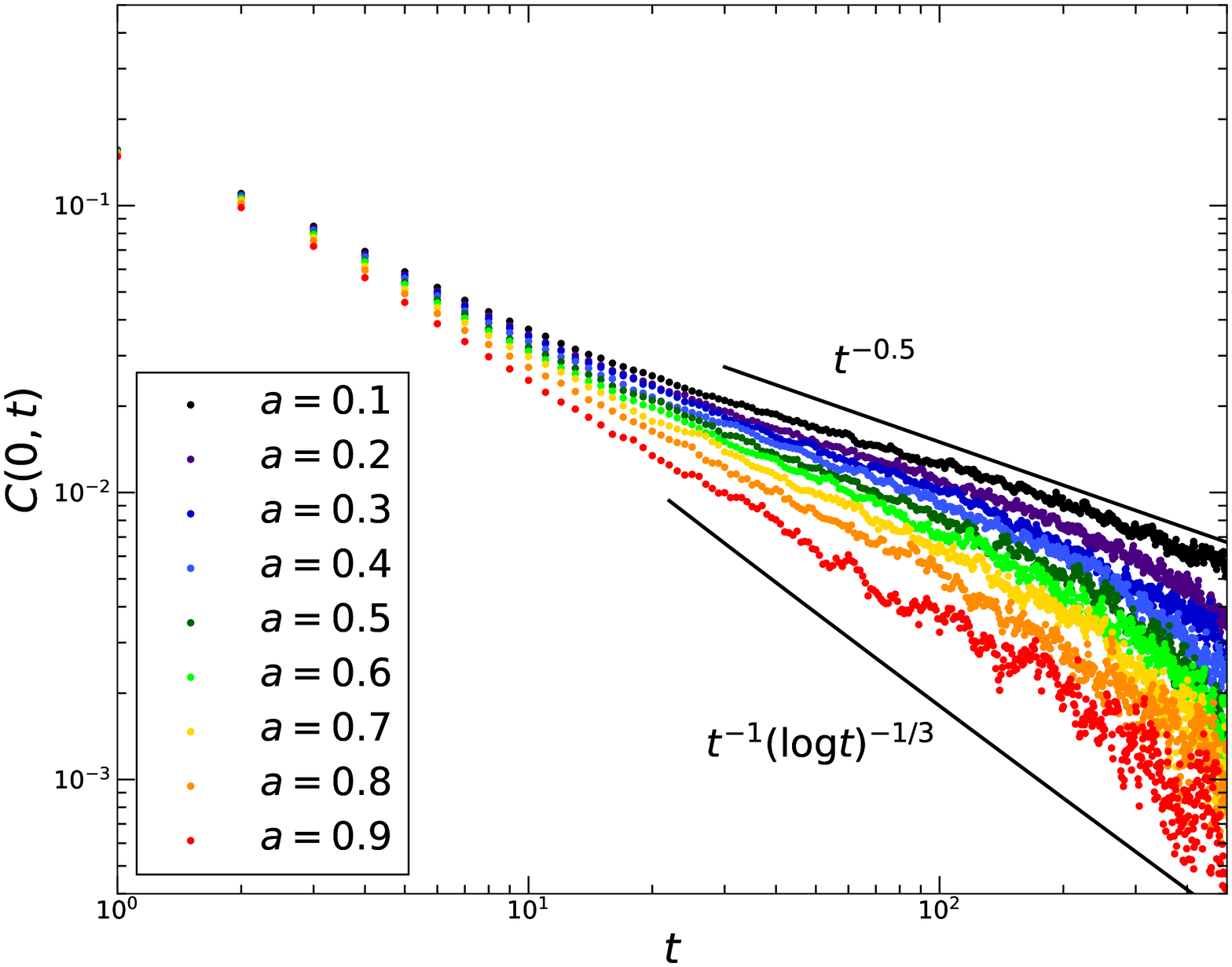}}
\hfill
\subfigure[\label{fig:int_fluct_size}]{\includegraphics[width=\columnwidth, trim={0cm 0cm 0cm 0cm}]{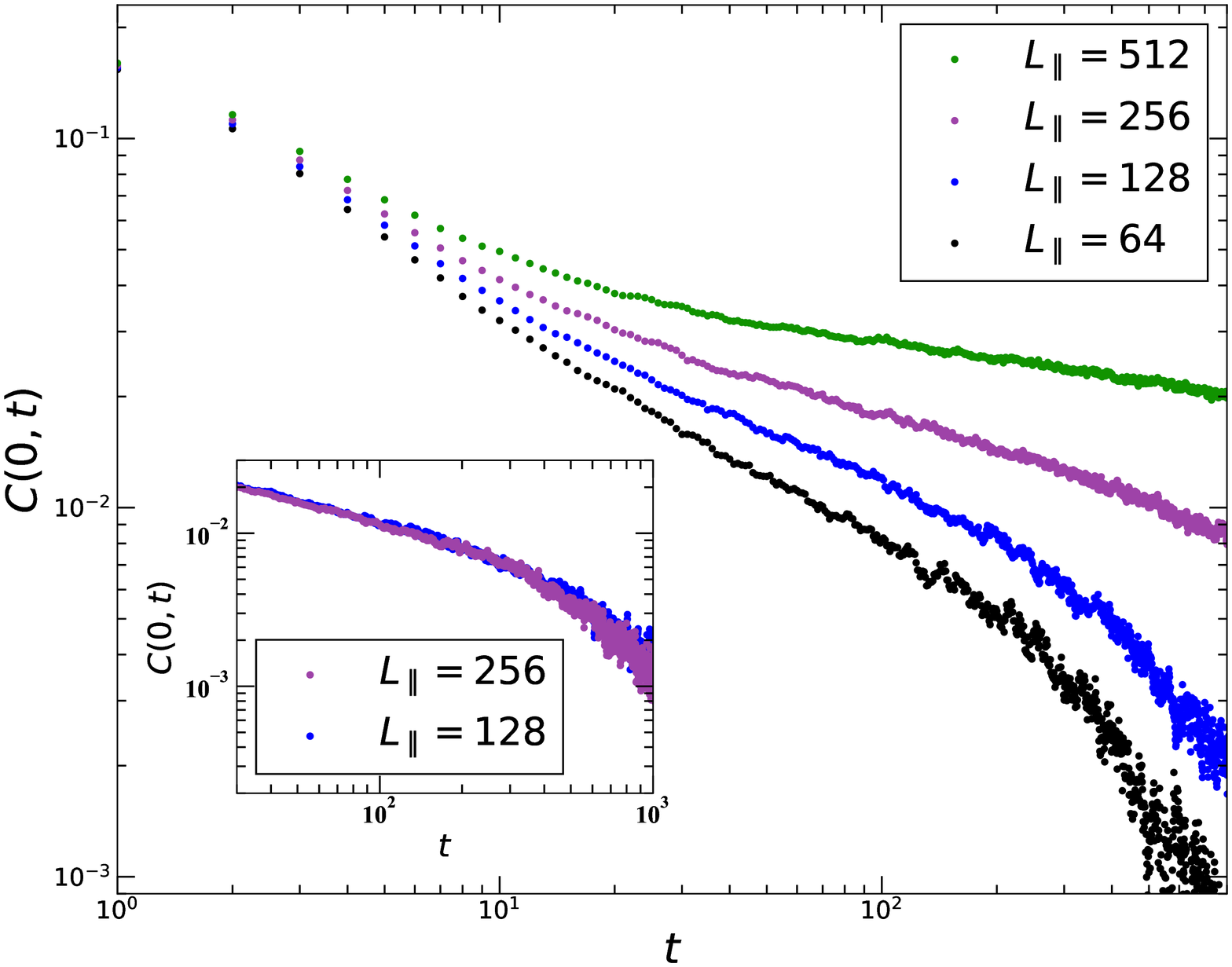}}
\caption{\label{fig:int_fluct} Double-logarithmic plots of the steady-state auto-correlation function decay with time at the phase interface within the hot subsystem of the two-temperature KLS driven lattice gas: 
(a) for different aspect ratios $a$ at fixed lattice dimensions $L_\parallel = 64$ and $L_\perp=32$;  
(b) for different total system lengths $L_\parallel$ at fixed width $L_\perp = 32$ and hot-to-critical subsystem ratio 1:1 ($a = 0.5$).
The inset compares two systems with different $L_\parallel$ and aspect ratios $a$, but with the same mean steady-state particle current $\langle J_{\parallel, st}\rangle$. 
All data shown pertain to $T_h = 2.0$ and $T_l = 0.8 \approx T_c$.
In all graphs, each data point was averaged over $50,000$ realizations.}
\end{figure*}
\subsection{\label{sec:sublevel53}Phase Interface Fluctuations}

Certain features of the two-temperature KLS driven lattice gas make studying the phase interface fluctuations within the hot subsystem with conventional means problematic.
First, instead of two well-separated phases of particles and holes, the hot subsystem is comprised of high- and low-density phases, which renders the task of tracing the phase interface ambiguous and rather arduous.  
Moreover, we cannot determine the exact time when the phase interface is fully formed. 
At the time instant when the two density shock waves meet at the middle of the hot subsystem, the particles at the center of the region are still uniformly distributed; only a few hundred Monte Carlo steps later the interface becomes fully established.  

To circumvent the outlined obstacles, we have decided to monitor the temporal decay of the steady-state auto-correlation function from the time instant $t_{ss}$ when the mean total particle current in the system does not change anymore,
\begin{align}\label{eqn:int_fluct}
\begin{split}
&C(\vec{x}=0;\, t - t_{ss}) =\\
&\ =\frac{1}{L_\perp} \sum_{j=1}^{L_\perp} \left[ \langle n_{x^*,j}(t) n_{x^*, j}(t_{ss}) \rangle - \langle n_{x^*,j}(t) \rangle  \langle n_{x^*,j}(t_{ss}) \rangle \right] ,
\end{split}
\end{align} 
where $n$ represents the occupancy of the site, and $x^* = a L_\parallel / 2$ is the location of the column in the center of the hot subsystem around which the phase interface forms. 
This steady-state density auto-correlation function reflects the dynamics of inteface fluctuations:
Comparing successively taken snapshots of the center column indicates how fast the phase interface shape varies with time. 

Asymptotically, both the (T)ASEP and KLS auto-correlations decay algebraically $C(0,t) \sim t^{- \zeta}$ as $t \to \infty$, where in two dimensions $\zeta = 1$ for the (T)ASEP, and $\zeta = 1/2$ for the critical KLS model, see Table~\ref{tab:table}.
More precisely, the TASEP auto-correlations display logarithmic corrections to the mean-field power law at the upper critical dimension $d_c = 2$, $C(0,0,t) \sim [t (\log t )^{1/3}]^{-1}$ \cite{Janssen:1986TASEP}.
Hence we expect the following range for any effective, i.e., potentially still size- and time-dependent auto-correlation decay exponent: $1/2 \leq \hat{\zeta} < 1$.
When both temperatures of our inhomogeneous KLS system are set far above the critical temperature, it indeed behaves like a two-dimensional (T)ASEP, with the auto-correlation function \eqref{eqn:int_fluct} decaying essentially linearly with time irrespective of the aspect ratio.

Yet if we set $T_l = 0.8 \approx T_c$, we have found after extensive analysis of our Monte Carlo simulation data that the effective scaling exponent $\hat{\zeta}$ which can be extracted from the density auto-correlations indeed appears to take any value in the allowed range, depending on the simulation domain's aspect ratio $a$ and total system length $L_\parallel$, as demonstrated in Fig.~\ref{fig:int_fluct_spectrum}, with the critical KLS value $\hat{\zeta} \to 1/2$ as $a \to 0$, whereas the two-dimensional (T)ASEP scaling is approached for $a \to 1$. 
Interestingly, our simulation data appear to correctly capture the subtle logarithmic corrections in this limit.
Furthermore we observe that upon increasing the system length $L_\parallel$, the auto-correlation decay slows down, presumably owing to increasing influence of the critical fluctuations in the cooler subsystem.
In the thermodynamic limit, perhaps the universal critical KLS decay exponent $\zeta = 1/2$ might be reached, but likely only after a prohibitely long crossover period.

Upon varying the system parameters, we have discovered that the effective auto-correlation decay exponent $\hat{\zeta}$ appears to be controlled by the mean steady-state particle current $\langle J_{\parallel, st} \rangle$ in the system. 
To demonstrate that, we plot the auto-correlation function in Fig.~\ref{fig:int_fluct_size} for a few systems with different lengths $L_\parallel$, keeping the hot-to-critical subsystem ratio constant. 
As we showed in Fig.~\ref{fig:SteadCur}, $\langle J_{\parallel, st} \rangle$ decreases with the system's length if the other parameters remain unchanged; in addition, as the mean steady-state current in the system decreases, so does the auto-correlation decay slope. 
However, as depicted in the inset of Fig.~\ref{fig:int_fluct_size}, two different system geometries that share the same value of the mean steady-state current yield overlapping auto-correlation curves.

\section{\label{sec:level6}Concluding Remarks}

As we have elucidated above, the intriguing stationary-state as well as transient kinetics of non-equilibrium systems displaying generic scale invariance may become even more complex upon combining different models subject to distinct microscopic dynamical rules. 
In this work, we have shown that a spatially inhomogeneous KLS driven lattice gas with temperature interfaces generated perpendicular to the non-equilibrium drive and net particle current spontaneously produces spatial patterns similar to those observed in TASEP systems with open boundaries. 
Indeed, although in the fully biased (infinite-drive) limit the distinct temperatures only affect hops transverse to the drive, the reduced stationary current in the cooler region system triggers a transport blockage at the interface where particles try to leave the hot subsystem.
As a result, the hot subsystem experiences phase separation, which destroys generic scale invariance in that region, a truly drastic and unexpected boundary effect.

When part of the lattice is maintained at the critical temperature $T_c$ for phase ordering, while the other subsystem is held at $T > T_c$, we observe imprints of the critical cluster fluctuations on both temperature regions:
(i) Near both interfaces, the critical region displays algebraic density decay.
(ii) The strong critical correlations that span across the entire system induce marked corrections relative to the mean-field predictions to the detailed shape of the density profile and in the high-temperature region as well as on the dynamics of the interface fluctuations, which appear to be controlled solely by the value of the stationary-state current.

Our specific choice of geometry for the two subsystems held at different temperatures created an inhomogeneous system wherein generic scale invariance becomes destroyed in the hotter region, while pertinent critical features in the region held at $T_c$ persist. 
We expect to observe drastically different dynamical features and large-scale properties when the temperature interfaces in the KLS driven lattice gas are oriented parallel to the drive. 
We plan to investigate that system's properties in the near future, as well as generalizations of both model variants in higher spatial dimensions.
\\

\begin{acknowledgments}
We thank Bart Brown, Michel Pleimling, and Royce Zia for fruitful discussions.
Research was sponsored by the Army Research Office and was accomplished under Grant Number W911NF-17-1-0156. 
The views and conclusions contained in this document are those of the authors and should not be interpreted as representing the official policies, either expressed or implied, of the Army Research Office or the U.S. Government. 
The U.S. Government is authorized to reproduce and distribute reprints for Government purposes notwithstanding any copyright notation herein. 
\end{acknowledgments}

\bibliography{mybib_no_links}

\end{document}